\documentclass[aps,prl,twocolumn,groupedaddress,showpacs,citeautoscript]{revtex4-1}

\usepackage{mathrsfs}
\usepackage{amsmath}
\usepackage{amssymb}
\usepackage{bm}
\usepackage{graphicx}
\usepackage{color}

\begin{document}


\title{Nonlinear optical conductivity of $U(1)$ spin liquids with large spinon Fermi surfaces}


\author{Yuan-Fei Ma}
\author{Tai-Kai Ng}
\affiliation{Department of Physics, Hong Kong University of Science and Technology, Clear Water Bay Road, Kowloon, Hong Kong, China}


\date{\today}

\begin{abstract}
In this paper we study the nonlinear current response of $U(1)$ spin liquids with large spinon Fermi surfaces under the perturbation of a time-dependent ac electric field $\mathbf{E}(t)$ within the framework of an effective $U(1)$ gauge theory. In particular, the third-order nonlinear current response to ac electric fields is derived. We show that as in the case of linear current response, an in-gap power-law ($\sim\omega^{\eta}$) response is found for the nonlinear current at low frequency. The nonlinear susceptibility may also induce through process of third harmonic generation propagating EM wave with frequency $3\omega$ inside the spin liquids.
\end{abstract}


\maketitle
\allowdisplaybreaks
\newcommand{\ud}{\mathrm{d}}


A Quantum spin liquid state is a Mott insulator without any long-range magnetic order down to the lowest temperatures \cite{Lee2008b, Balents2010}. For a long time, this exotic state exists only in the theorists' models. Recent experimental progress on the organic compound $\kappa$-${\rm (ET)_2Cu_2(CN)_3}$ \cite{Shimizu2003, Kezsmarki2006, Kurosaki2005, Yamashita2008a, Yamashita2008} and a few other candidate materials \cite{Helton2007, Okamoto2007, Zhou2011, Itou2008, Han2012} have changed the situation. In the case of $\kappa$-${\rm (ET)_2Cu_2(CN)_3}$, the system is described by a nearly isotropic spin-$1/2$ Hubbard model on the triangular lattice, and it is believed that the insulator state is near the Mott transition, where the charge excitations are gapped and the spin-$1/2$ spinons form a Fermi surface. The dynamics of the spin liquid state is described by an effective $U(1)$ gauge theory \cite{Lee2005, Ng2007}. Several experimental results have confirmed the Fermi-liquid-like behavior like the linear temperature dependence of the specific heat, the Pauli-like spin susceptibility, the nearly unity Wilson ratio, etc.. These experiments suggest a deep relationship between the unfamiliar spin liquid state and the well-studied Fermi liquids \cite{Zhou2013a}.

In the continuum limit, all existing theoretical studies suggested that the spin liquids can be described by an effective $U(1)$ gauge theory Lagrangian:
\begin{eqnarray}
\label{eq:Lagrangian}
 \mathcal{L}_{\rm eff} &=& \mathcal{L}(\psi^{\dag}_{\sigma}, \psi_{\sigma})  + \mathcal{L}_g(a_0, \mathbf{a}),  \\
 \mathcal{L} &=& \psi^{\dag}_{\sigma} \left(i\partial_t - a_0 - A_0 \right)\psi_{\sigma} - \frac{1}{2m_0} |(\nabla - i\mathbf{a} - i\mathbf{A})\psi_{\sigma}|^2, \nonumber \\
 \mathcal{L}_g &=& \frac{1}{2g}\left[ \frac{1}{v^2_a}\left( \frac{\partial \mathbf{a}}{\partial t}\right)^2- (\nabla \times \mathbf{a})^2\right] + \mathcal{L}_l^{(2)}(a_0) + \mathcal{L}_g^{(4)}(a_0, \mathbf{a}). \nonumber
\end{eqnarray}
Here $\psi^{\dag}_{\sigma}$($\psi_{\sigma}$) are spin-$\sigma$ fermion creation (annihilation) operators, $(A_0, \mathbf{A})$ is the external applied electromagnetic field, and $g$ and $v_a$ are the coupling constant and effective velocity of the internal gauge field, respectively. $\mathcal{L}_l^{(2)}$ is the effective (quadratic) action for the longitudinal gauge field $a_0$. The dynamics of the internal gauge field arises from integrating out the charge degree of the system through for example, the slave-rotor approach \cite{Lee2005} or from an effective Landau Fermi-liquid-like approach \cite{Zhou2013a}, typically to second order in $f_{\mu\nu} = \partial_{\mu} a_{\nu} - \partial_{\nu} a_{\mu}$. In this paper we keep also the $4^{\rm th}$ order terms in $\mathcal{L}_g$,
\[
\mathcal{L}_g^{(4)}(a_0, \mathbf{a})\sim b_1(\mathbf{e}\cdot\mathbf{e})^2 + b_2(\mathbf{b}\cdot\mathbf{b})^2 + b_3(\mathbf{e}\cdot\mathbf{e})(\mathbf{b}\cdot\mathbf{b}) + b_4(\mathbf{e}\cdot\mathbf{b})^2,
\]
where $\mathbf{e}=- \partial_t\mathbf{a}-\nabla a_0$ and $\mathbf{b}=\nabla\times\mathbf{a}$ are the internal electric and magnetic fields associated with the internal gauge field $a_{\mu}$, respectively. These terms arises naturally in a lattice gauge theory model \cite{Lee2005}. We have assumed that the spin liquid system respects time-reversal, inversion and rotational symmetry in writing down $\mathcal{L}_g^{(4)}$. The dynamics of internal gauge field are all expressed in terms of the field strengths $\mathbf{e}$ and $\mathbf{b}$ because of gauge invariance associated with internal gauge field in the insulator state \cite{Lee2005}.

Previous studies in the electrodynamic properties of spin liquid states mainly focused on the linear response of the system, i.e. the conductivity tensor and dielectric function. In this paper we study the nonlinear response of $U(1)$ spin liquids with large Fermi surface to external EM fields. We shall focus on the $\mathbf{q}\rightarrow0, \omega\neq 0$ responses in this paper and shall derive the third-order electric conductivity $\sigma^{(3)}(\omega)$ for the spin liquid state. In particular, we are interested at examining whether a non-linear $I-V$ relation $I\sim V^{1+\beta} (\beta>0)$ may exist in spin liquid states. This is driven by the predicted existence of power-law linear conductivity $\sigma^{(1)}(\omega)\sim\omega^{2+\alpha}$ in $U(1)$ spin liquids. The power-law conductivity leads to nonlinear $I-V$ relation in many one-dimensional Luttinger liquid systems \cite{giuliani2005}.

Our approach consists of two steps: first we derive an equation of motion for the fermion (spinon) current under the perturbation of a time-dependent but spatially uniform electric field $\mathbf{E}(t)$ from which the nonlinear responses of the spinons to the total electric field can be computed. The equation of motion is supplemented by an equation determining the internal electric field $\mathbf{e}$ and the two equations of motion are solved self-consistently to determine the nonlinear responses of the spin liquids to external electric fields.

\subsection{Equation of motion for fermion (spinon) current}
We first consider the equation of motion for the fermion current under a time-dependent but spatially uniform electric field. In this case it is most convenient to perform a transformation to a non-inertial frame which moves with the center of the mass of the spin liquid system \cite{Ng2005}. In this coordinate system, the time-dependent center-of-mass position vector $\mathbf{R}(t) \equiv 1/N \sum_i \langle \mathbf{r}_i(t)\rangle$ ($N=$ total number of fermions) satisfies the equation of motion:
\begin{eqnarray}
  m_0 \ddot{\mathbf{R}}(t)  &=& \frac{1}{N} \int \ud^d \mathbf{r} \, n(\mathbf{r}, t) \mathbf{\mathcal{E}}(\mathbf{r}, t) -\frac{1}{N} \int \ud^d \mathbf{r} \, n(\mathbf{r}, t) U(\mathbf{r}) \nonumber \\
  && + \frac{1}{N} \int \ud^d \mathbf{r} \, \mathbf{j}(\mathbf{r}, t) \times \mathbf{\mathcal{B}}(\mathbf{r}, t),
  \label{eq:eom}
\end{eqnarray}
where $n(\mathbf{r}, t) = \langle \sum_i \delta (\mathbf{r} - \mathbf{r}_i) \rangle$ is the time dependent particle density distribution function. And
\[ \mathbf{j}(\mathbf{r}, t) = \mathbf{j}_0(\mathbf{r},t) - \frac{n(\mathbf{r}, t)}{m_0}(\mathbf{A}(\mathbf{r}, t) + \mathbf{a}(\mathbf{r}, t))
\]
is the total current density where $\mathbf{j}_0(\mathbf{r},t) = \langle \sum_i [\frac{\mathbf{p}_i}{2m_0} \delta (\mathbf{r} - \mathbf{r}_i) + \delta (\mathbf{r} - \mathbf{r}_i) \frac{\mathbf{p}_i}{2m_0} ]\rangle$ is the canonical current density, $\mathbf{p}=$ canonical momentum. $\mathbf{\mathcal{E}}=\mathbf{E}+\mathbf{e}$ and $\mathbf{\mathcal{B}}=\mathbf{B}+
\mathbf{b}$ are the total electric and magnetic fields ($\mathbf{E}$ and $\mathbf{B}$ are the external electric and magnetic fields, respectively), and $U(\mathbf{r})$ is a background disordered scalar potential the fermions see. The meaning of this equation is clear: the first term on the right hand side is just the electric force, the second term comes from disordered potential and the last term corresponds to the Lorentz force contribution.

The disordered potential $U(\mathbf{r}) = \lambda u(\mathbf{r})$ satisfies:
 \begin{eqnarray}
 \langle u(\mathbf{r}) \rangle =0, \, \langle u(\mathbf{r}) u(\mathbf{r}') \rangle = |u|^2 \delta(\mathbf{r} - \mathbf{r}'),
 \end{eqnarray}
and can be integrated out to second term in $\lambda^2$ \cite{Ng2005}:
 \begin{eqnarray}
  &&-\frac{1}{N} \int \ud^d \mathbf{r} \, n(\mathbf{r}, t) U(\mathbf{r}) \nonumber \\
   &\simeq& \frac{\lambda^2 |u|^2}{n}  \int \ud t' \,\, \nabla_{\mathbf{R}(t)} \chi_d(\mathbf{R}(t)-\mathbf{R}(t'), t-t'),
   \label{eq:ddresponse}
\end{eqnarray}
where $\chi_d(\mathbf{r}- \mathbf{r}', t-t')$ is the density-density response function for the system in the absence of impurities.

The equation can be simplified further if the applied electric field $\mathbf{E}(t)\sim\mathbf{E}_0 e^{-i\omega t}$ is weak and it's frequency $\omega$ is small. In this limit $\mathbf{R}(t)$ is slowly varying and we may approximate $\mathbf{R}(t)-\mathbf{R}(t') \sim \dot{\mathbf{R}}(t)  (t - t') + (1/2)\ddot{\mathbf{R}}(t)  (t - t')^2 $ and neglect higher order time-derivative terms. Then for a system with inversion symmetry $\chi_d(-\mathbf{q}, \omega) = \chi_d(\mathbf{q}, \omega)$, the right hand side of Eq.(\ref{eq:ddresponse}) is related to the density-density response function through \cite{Ng2005}:
\begin{eqnarray}
  &&\int \ud t' \,\, \nabla_{\mathbf{R}(t)} \chi_d(\mathbf{R}(t)-\mathbf{R}(t'), t-t') \nonumber \\
   &=& \sum_{\mathbf{q}} i \mathbf{q} \left(1 - \frac{i}{2} \mathbf{q} \cdot \ddot{\mathbf{R}}(t) \frac{\partial^2}{\partial \omega'^2} \right)\chi_d(\mathbf{q}, \omega')|_{\omega' = \mathbf{q} \cdot \dot{\mathbf{R}}(t)}.
\label{eq:dd2}
\end{eqnarray}
Putting together Eqs.\ (\ref{eq:eom}),\ (\ref{eq:ddresponse}) and\ (\ref{eq:dd2}), we obtain the equation of motion:
\begin{eqnarray}
  &&m\cdot \ddot{\mathbf{R}}(t)  = \frac{1}{N} \int \ud^d \mathbf{r} \, n(\mathbf{r}, t) \mathbf{\mathcal{E}}(\mathbf{r}, t) \\
  && - \frac{\lambda^2 |u|^2}{n} \sum_{\mathbf{q}} \mathbf{q}\, \mathrm{Im}\chi_d(\mathbf{q}, \mathbf{q} \cdot \dot{\mathbf{R}}(t)) + \frac{1}{N} \int \ud^d \mathbf{r} \, \mathbf{j}(\mathbf{r}, t) \times \mathbf{\mathcal{B}}(\mathbf{r}, t), \nonumber
\end{eqnarray}
where the effective mass is given by:
\begin{eqnarray}
  m = m_0 - \frac{\lambda^2 |u|^2}{2n}\sum_{\mathbf{q}} \frac{q^2}{d} \frac{\partial^2}{\partial \omega'^2} \mathrm{Re} \chi_d(\mathbf{q}, \omega')|_{\omega' = \mathbf{q} \cdot \dot{\mathbf{R}}(t)},
\end{eqnarray}
where $d=$ system dimension. Notice that the second order term $\ddot{\mathbf{R}}(t)  (t - t')^2$ only leads to a mass renormalization that is proportional to the strength of disorder.

As $\mathrm{Im} \chi_d(\mathbf{q}, \omega)\neq0$ only at $|\omega/q v_F|< 1$, the leading order contributions to the imaginary part of the density-density response function is given by:
\begin{eqnarray}
  \mathrm{Im} \chi_d(\mathbf{q}, \omega) \sim \alpha_1(q) \frac{\omega}{q v_F} + \alpha_2(q) \left(\frac{\omega}{q v_F}\right)^3 + \cdots,
\end{eqnarray}
where even order terms are zero, $\alpha_1(q)$ and $\alpha_2(q)$ are $\omega$-independent parameters. In 2D, we obtain
\begin{eqnarray}
  \alpha_1(q) &=& N(0) \left[1 + \frac{1}{2} \left(\frac{q}{2k_F}\right)^2 + O((q/2k_F)^4) \right], \nonumber \\
  \alpha_2(q) &=& N(0) \left[\frac{1}{2} + \frac{5}{4} \left(\frac{q}{2k_F}\right)^3 + O((q/2k_F)^5) \right]. \nonumber
\end{eqnarray}

Thus we obtain
\begin{eqnarray}
  &&\frac{\lambda^2 |u|^2}{n} \int \ud t' \,\, \nabla_{\mathbf{R}(t)} \chi_d(\mathbf{R}(t)-\mathbf{R}(t'), t-t') \nonumber \\
  &=& -\frac{m}{\tau} \dot{\mathbf{R}}(t) + \gamma m^3 \dot{\mathbf{R}}(t) \dot{\mathbf{R}}(t)^2,
\end{eqnarray}
in the small $\omega$ limit where $\tau$ and $\gamma$ are related to $\alpha_1(q)$ and $\alpha_2(q)$ by
\begin{eqnarray}
  \frac{1}{\tau} &=& \frac{\lambda^2|u|^2}{n k_F} \int \ud q \ud \Omega\, q^d \alpha_1(q) \cos^2{\theta}, \nonumber \\
  \gamma &=& -\frac{\lambda^2|u|^2}{n k^3_F} \int \ud q \ud \Omega\, q^d \alpha_2(q) \cos^4{\theta}.
  \label{eq:taugam}
\end{eqnarray}

Therefore in the absence of external magnetic field, the equation of motion for the charge current $\mathbf{j}(t) \equiv n_0 e \dot{\mathbf{R}}(t)$ in the weak field limit is:
\begin{eqnarray}
  &&m \dot{\mathbf{j}}(t) = n_0 e^2 \mathbf{E}(t) - \frac{m}{\tau} \mathbf{j}(t) +  n_0 e^2 \int \ud^d \mathbf{r}\,[n_0 + \delta n(\mathbf{r},t)] \mathbf{e}(\mathbf{r},t) \nonumber \\
  && + \frac{\gamma m^3}{(n_0 e)^2} \mathbf{j}(t) [\mathbf{j}(t) \cdot \mathbf{j}(t)]  + e\int \ud^d \mathbf{r}\, \mathbf{j}(\mathbf{r}, t) \times \mathbf{b}(\mathbf{r}, t),
  \label{eq:eqcurrent}
\end{eqnarray}
where the particle density has been decomposed into a constant part $n_0$ and a varying part $\delta n(\mathbf{r},t)$. In the $\mathbf{q}\rightarrow0$ limit, $\delta n(\mathbf{r},t), \mathbf{b}(\mathbf{r}, t)\rightarrow0$ and the equation reproduces the usual Drude conductivity if the non-linear $\mathbf{j}^3$ term and the internal electric field $\mathbf{e}=-\partial_t \mathbf{a} - \nabla a_0$ are neglected.

\subsection{Third order nonlinear current response}
To solve the above equation we need to determine the internal gauge field $\mathbf{a}$. The internal gauge field $\mathbf{a}$ can be determined by the Lagrangian equation of motion $\delta\mathcal{L}_{\rm eff}/\delta\mathbf{a}=0$, which leads to the equation of motion (in the Coulomb gauge and in the limit $\mathbf{q}\rightarrow0$):
\begin{equation}
\label{eqma}
\frac{1}{v_a^{2}} \frac{\partial^2}{\partial t^2} \mathbf{a}(\mathbf{r}, t) + 4 b_1 \frac{\partial}{\partial t} \left[ \left(\partial_t{\mathbf{a}}(\mathbf{r},t)\right)^2 \partial_t{\mathbf{a}}(\mathbf{r},t) \right] = g \mathbf{j}(\mathbf{r}, t).
\end{equation}

Eqs.~(\ref{eq:eqcurrent}) and (\ref{eqma}) together determine the current response under the time-dependent electric field $\mathbf{E}(t)$, which contain nonlinear effects up to third order in $\mathbf{E}$. To solve the equations, we first Fourier transform Eq.\ (\ref{eqma}) in the limit $\mathbf{q}\rightarrow0$,
\[
    \frac{i\omega}{v^2_a}\mathbf{e}(\omega) + 4 b_1 i\omega \sum_{(lmn)} \mathbf{e}(\omega_l)[\mathbf{e}(\omega_m) \cdot \mathbf{e}(\omega_n)] = g\mathbf{j}(\omega),
\]
where $\sum_{(lmn)} \equiv \sum_{\omega_l, \omega_m, \omega_n} \delta(\omega-\omega_l-\omega_m-\omega_n)$.

Therefore up to third order, the internal electric field $\mathbf{e}(\omega)$ is related to the total current $\mathbf{j}(\omega)$ by
\[
\mathbf{e}(\omega)=\frac{g v^2_a}{i\omega} \mathbf{j}(\omega) + i b \sum_{(lmn)} \frac{ \mathbf{j}(\omega_l)[\mathbf{j}(\omega_m) \cdot \mathbf{j}(\omega_n)]}{\omega_l \omega_m \omega_n},
\]
where $b=4b_1g^3v^8_a$.

Lastly, we insert this result into Eq.\ (\ref{eq:eqcurrent}), obtaining
\begin{eqnarray}
&&m (-i\omega) {\mathbf{j}}(\omega) = n_0 e^2 \mathbf{E}(\omega)+n_0 e^2 gv^2_a \frac{\mathbf{j}(\omega)}{i\omega} - \frac{m}{\tau} \mathbf{j}(\omega) \nonumber \\
&&+ \frac{\gamma m^3}{(n_0 e)^2} \sum_{(lmn)} \left[1 + \frac{i \bar{b}}{\omega_l \omega_m \omega_n} \right] \mathbf{j}(\omega_l) [\mathbf{j}(\omega_m) \cdot \mathbf{j}(\omega_n)]. \label{eq:eqcurrent2}
\end{eqnarray}
where $\bar{b} = n^3_0 e^4 b/ \gamma m^3$.

It is easy to see that up to third order in the external electric field $\mathbf{E}$ the total current is given by:
\begin{eqnarray}
  \mathbf{j}(\omega) = \sigma^{(1)} (\omega) \mathbf{E}(\omega) + \bar{\gamma} \sigma^{(1)} (\omega) \sum_{(lmn)} \left[1 + \frac{i \bar{b}}{\omega_l \omega_m \omega_n} \right] \times \nonumber \\
   \sigma^{(1)}(\omega_l) \sigma^{(1)}(\omega_m) \sigma^{(1)}(\omega_n)
   \mathbf{E}(\omega_l)\left(\mathbf{E}(\omega_m) \cdot \mathbf{E}(\omega_n)\right), \nonumber \\
\end{eqnarray}
where $\bar{\gamma} = \gamma m^3/n^3_0 e^4$ and $\sigma^{(1)} (\omega)$ is the linear conductivity given by:
\begin{eqnarray}
  \sigma^{(1)} (\omega) = \frac{\omega \sigma_0(\omega)}{\omega + i \beta \sigma_0(\omega)},
\end{eqnarray}
where $\sigma_0(\omega)$ is the ac Drude conductivity $\sigma_0(\omega) = \sigma_0/(1 - i \omega \tau)$ and $\beta=gv^2_a$. We note that there is no distinction between longitudinal and transverse currents in the limit $\mathbf{q}\rightarrow0$. The linear conductivity $\sigma^{(1)} (\omega)$ is consistent with the result obtained through Ioffe-Larkin composition rule in the framework of $U(1)$ gauge theory \cite{Ng2007, Zhou2013a}, which leads to the power-law optical conductivity in the Mott gap. The nonlinear terms are results of non-linearity in the quasi-particle scattering ($\mathrm{Im}\chi_d$) and the presence of $\mathcal{L}_g^{(4)}$. In the small-$\omega$ limit, $ \sigma^{(1)} (\omega) \sim \omega/(i \beta)$, and the nonlinear response reduces to a simple form:
\begin{eqnarray}
 \mathbf{j}^{(3)}(\omega)\sim-\frac{\bar{\gamma}\bar{b}}{\beta^3} \sigma^{(1)}(\omega)\sum_{(lmn)}
 \mathbf{E}(\omega_l)\left( \mathbf{E}(\omega_m) \cdot \mathbf{E}(\omega_n)\right).
 \label{eq:nonlinearj3}
\end{eqnarray}

Therefore our theory predicts that in the low frequency limit, there is a third-order correction to the current, i.e.,
\begin{eqnarray}
  &&\mathbf{j}(\omega) = \sigma^{(1)} (\omega)\mathbf{E}(\omega) \nonumber \\
  &&+ \sum_{(lmn)} \sigma^{(3)}(\omega_l+\omega_m+\omega_n)\mathbf{E}(\omega_l) \left(\mathbf{E}(\omega_m) \cdot \mathbf{E}(\omega_n)\right),
\end{eqnarray}
with $\sigma^{(3)}(\omega)$ proportional to $\sigma^{(1)}(\omega)$. In particular, we do not find any nonlinear $I-V$ relation.

\subsection{Nonlinear optical wave generation}
The nonlinear conductivity leads to another interesting consequence when the optical response of the system is considered. We note that the spin liquid system can be treated as a nonlinear medium, where the linear electric susceptibility $\chi^{(1)}$ and the third-order nonlinear electric susceptibility $\chi^{(3)}$ are given by:
\begin{subequations}
\begin{eqnarray}
  \chi^{(1)}(\omega; \omega_n) &=& \frac{i \sigma_0(\omega)}{ \omega + i \beta \sigma_0(\omega)}, \\
  \chi^{(3)}(\omega; \omega_l, \omega_m, \omega_n) &=& \bar{\gamma} \chi^{(1)} (\omega) \left[1 + \frac{i \bar{b}}{\omega_l \omega_m \omega_n} \right] \times \nonumber \\
  &&\sigma^{(1)}(\omega_l) \sigma^{(1)}(\omega_m) \sigma^{(1)}(\omega_n).
\end{eqnarray}
\end{subequations}

As shown in Ma and Ng \cite{Ma&Ng}, the complex linear susceptibility leads to surface plasmon modes propagating along the interface between a linear medium and the spin liquids in the linear response regime. Assuming a boundary along $y$-direction at $x=0$, the surface mode has a form $\mathbf{E}_S(\mathbf{r},t) = \mathbf{A}_1(\omega) e^{i \mathbf{k}_1 \cdot \mathbf{r} - i \omega t} + c.c.$, where $\mathbf{k}_1 = (k_{1s}, k, 0)$ is the wave vector, and $\mathbf{A}_1(\omega)$ is the amplitude of the electric field. The components of the wave vector $\mathbf{k}_1$ are given by \cite{raether1988surface}:
\begin{subequations}
\begin{eqnarray}
  k^2 &=& \frac{\omega^2}{c^2} \frac{\varepsilon_0 \varepsilon^{(1)}(\omega)}{\varepsilon_0 +  \varepsilon^{(1)}(\omega)},  \\
  k^2_{1s} &=& \frac{\omega^2}{c^2} \frac{[\varepsilon^{(1)}(\omega)]^2}{\varepsilon_0 + \varepsilon^{(1)}(\omega)},
\end{eqnarray}
\end{subequations}
where $\varepsilon_0$ is the dielectric constant of the linear medium (the vacuum), and $\varepsilon^{(1)}(\omega) = 1 + 4\pi \chi^{(1)}(\omega)$ is the dielectric function of the nonlinear medium (the spin liquids). We are interested at surface modes where $k$ is nearly real and $k_{1s}$ is nearly imaginary. It was shown in Ma and Ng \cite{Ma&Ng} that a sizable frequency range where this condition is met is found in spin liquids (with large spinon Fermi surfaces) close to metal-insulator transition.

The surface wave will induce a nonlinear electric polarization:
\begin{eqnarray}
  \mathbf{P}_3(3\omega) = \mathbf{e}_3 \chi^{(3)}(3\omega; \omega, \omega, \omega) A^3_1(\omega)e^{i\mathbf{k}_{3s} \cdot \mathbf{r} - i 3\omega t},
\end{eqnarray}
where $\mathbf{e}_3$ is the direction of the polarization vector, and $\mathbf{k}_{3s} = 3 \mathbf{k}_1$. The nonlinear polarization $\mathbf{P}_3(3\omega)$ acts as a source term in the nonlinear wave equation
\begin{subequations}
\begin{eqnarray}
  \nabla \times \nabla \times \mathbf{E}_s(3\omega) - \frac{(3\omega)^2}{c^2} \varepsilon^{(1)}(3\omega) \cdot \mathbf{E}_s(3\omega) \nonumber \\
  = \frac{4\pi (3\omega)^2}{c^2} \mathbf{e}_3 P_3(3\omega) e^{i\mathbf{k}_{3s} \cdot \mathbf{r}}.
\end{eqnarray}
and generates radiation with frequency $3\omega$ \cite{shen1984principles, boyd2008nonlinear}. The solution of the nonlinear wave equation contains a particular part with wave vector $\mathbf{k}_{3s}$ and a homogeneous part with wave vector $\mathbf{k}_3$ where $k^2_3 = \frac{(3\omega)^2}{c^2} \varepsilon^{(1)}(3\omega)$, i.e.,
\begin{eqnarray}
  \mathbf{E}_s(3\omega) = \mathbf{A}_{3P}(3\omega) e^{i \mathbf{k}_{3s} \cdot \mathbf{r}} + \mathbf{A}_{3H}(3\omega) e^{i \mathbf{k}_3 \cdot \mathbf{r}}.
\end{eqnarray}
\end{subequations}

\begin{figure}
\includegraphics[width = 0.45 \textwidth]{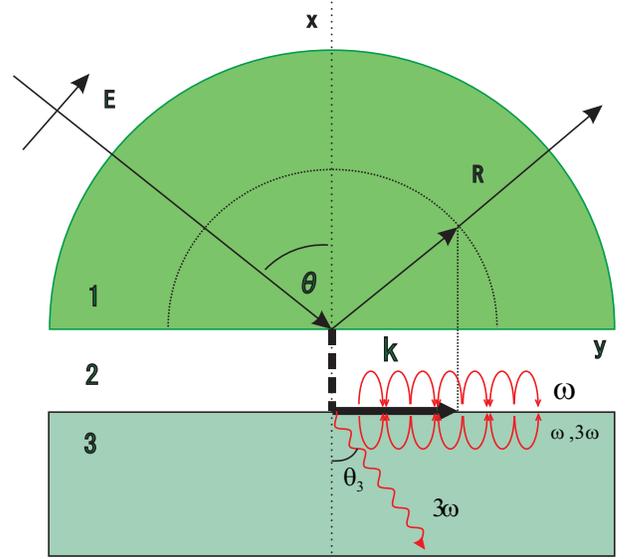}
\caption{\label{fig:nonlinear} The surface modes (with frequency $\omega$) supported by the linear complex susceptibility $\varepsilon^{(1)}$ (Here the Otto configuration has been used to excite the surface modes: the three layers ($1|2|3$) are prism, linear medium (e.g., the vacuum) and the spin liquids, respectively.) will induce the waves with frequency $3\omega$ that can propagate in the spin liquids. If the propagation angle $\theta_3\ll \pi/2$, it can be distinguished from other surface waves and is detectable.}
\end{figure}

In the presence of boundary, the particular and homogeneous solutions are coupled.
Solving the wave equations with an incoming wave with frequency $\omega$ at angle $\theta_I$ upon the boundary at $x=0$ (see Fig.~\ref{fig:nonlinear}), we find that the particular solution can be chosen as \cite{shen1984principles, boyd2008nonlinear}:
\begin{eqnarray}
  \mathbf{A}_{3P}(3\omega) = \frac{4\pi P_3(3\omega)}{\varepsilon_s(3\omega) - \varepsilon^{(1)}(3\omega)} \left(\mathbf{e}_3 - \frac{\mathbf{k}_{3s}(\mathbf{k}_{3s}\cdot\mathbf{e}_3)}{k^2_3} \right), \nonumber
\end{eqnarray}
where $\varepsilon_s(3\omega) \equiv \frac{c^2}{(3\omega)^2} k^2_{3s}$. Matching the phase factors at the boundary leads to Snell's law:
\begin{eqnarray}
  k_{3,y} = k_{3s, y} = 3k.
\end{eqnarray}
and the amplitude of the homogenous part is found to be:
\begin{eqnarray}
  A_{3H} = -\frac{4\pi P_3(3\omega)}{\varepsilon_s(3\omega) - \varepsilon^{(1)}(3\omega)} \frac{\sqrt{\varepsilon_0} \cos{\theta_{3s}} + \sqrt{\varepsilon_s(3\omega)} \cos{\theta_I}}{\sqrt{\varepsilon_0} \cos{\theta_{3}} + \sqrt{\varepsilon^{(1)}(3\omega)} \cos{\theta_I}}, \nonumber
\end{eqnarray}
where $\theta_{3s}$ and $\theta_3$ are the propagation directions of the particular part and the homogenous part, respectively. $\theta_{3s}$ is related to $\theta_I$ and $\omega$ through Snell's law:
\begin{eqnarray}
  \sqrt{\varepsilon_0} \sin{\theta_I} = \sqrt{\varepsilon^{(1)}(\omega)} \sin{\theta_{3s}}.
\end{eqnarray}

The particular part (with wavevector $\mathbf{k}_{3s} = 3 \mathbf{k}_1 =  (3 k_{1s}, 3k, 0)$) decays rapidly in the nonlinear medium as $k_{1s}$ is nearly pure imaginary when $k$ is real. However the homogeneous part decays rather slowly and can propagate in the nonlinear medium \cite{Ma&Ng}. The propagation direction of the waves is given by
\begin{eqnarray}
\label{angle}
  \tan{\theta_3} &=& \frac{k_{3,y}}{k_{3,x}} = \frac{3k}{\sqrt{\frac{(3\omega)^2}{c^2} \varepsilon^{(1)}(3\omega) - (3k)^2}}.
\end{eqnarray}
For $\theta_3 \ll\pi/2$, the wave will propagate away from the boundary and can be distinguished from the other surface waves as shown schematically in Fig.~\ref{fig:nonlinear}.

\begin{figure}
\includegraphics[width = 0.45 \textwidth]{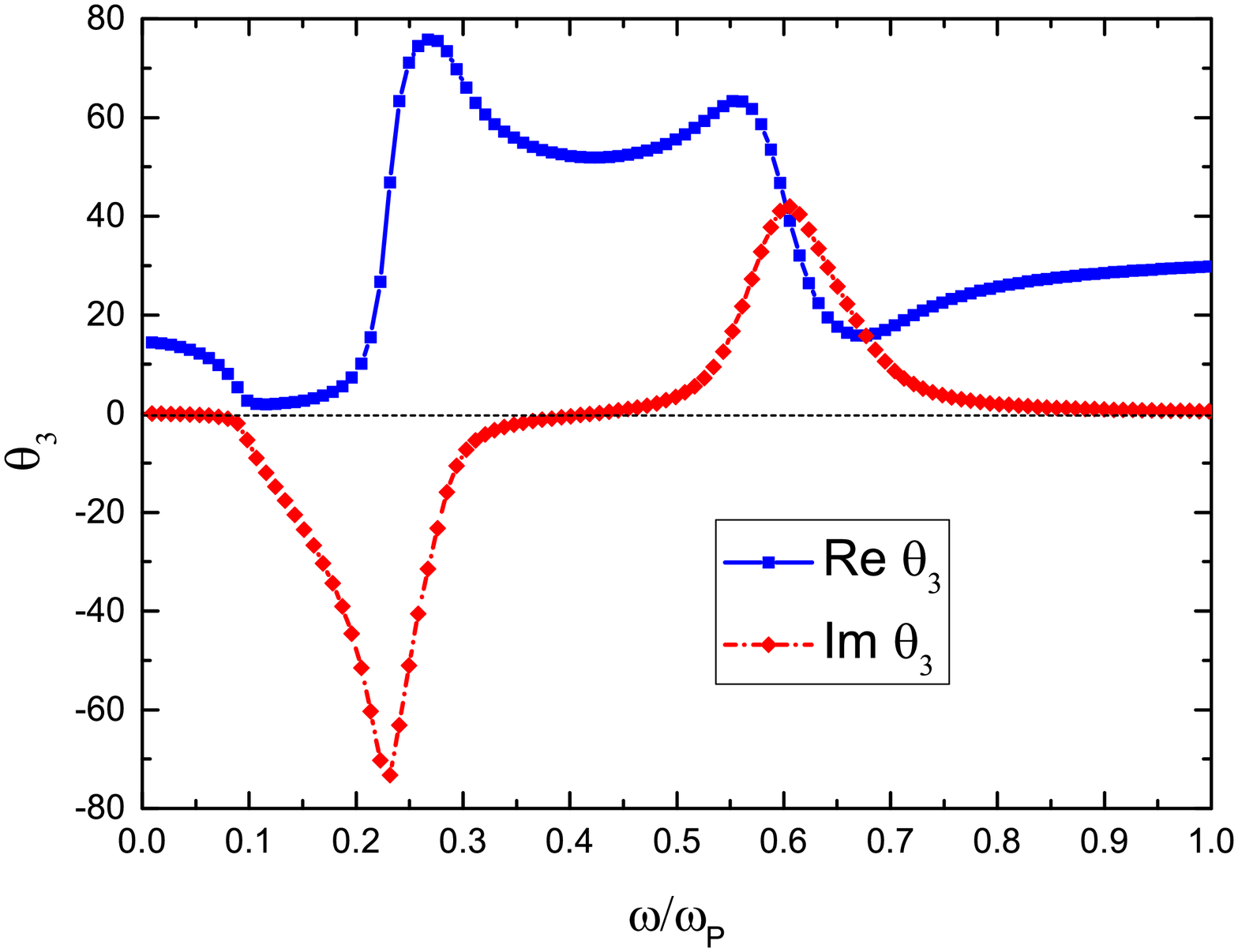}
\caption{\label{fig:theta3} Real and imaginary parts of the propagation angle $\theta_3$ of the waves with frequency $3\omega$. We take $\tau^{-1}=0.1 \omega$ in the calculation \cite{Ma&Ng}. It clearly shows that in some frequency region, the angle is nearly real $\theta_3 \ll \pi/2$. If we choose $\varepsilon(\infty) =2.5$, for incident wave with frequency $\omega = 0.4 \omega_P$, the propagation angle is about $\theta_3 \approx 52^\circ$.}
\end{figure}

For the $U(1)$ spin liquids, the linear dielectric function can be written as \cite{Ma&Ng},
\begin{eqnarray}
   \varepsilon^{(1)}(\omega) = \varepsilon(\infty) - \frac{\omega^2_P}{\omega^2 - \omega^2_g + i \omega \tau^{-1}},
\end{eqnarray}
where $\varepsilon(\infty)$ is the background dielectric constant, and $\omega^2_g = \beta \omega^2_P/4\pi$ with $\omega_P$ the plasma frequency. With Eq.~(\ref{angle}), it is straightforward to show that in the limit $\tau \rightarrow \infty$, $\theta_3<\pi/2$ if
\begin{eqnarray}
  \frac{1}{\varepsilon^{(1)}(3\omega)} - \frac{1}{\varepsilon^{(1)}(\omega)} < \frac{1}{\varepsilon_0},
\end{eqnarray}
where we choose $\varepsilon^{(1)}(3\omega) > 0$. For nonzero $\tau$ this condition is satisfied in a narrower frequency region. The real and imaginary parts of $\theta_3$ is shown in Fig.~\ref{fig:theta3} for $\tau=0.1\omega$ with a background dielectric constant $\varepsilon(\infty) =2.5$.
We see that there is a finite frequency range where ${\rm Im}\theta_3$ is small and ${\rm Re}\theta_3<\pi/2$. In particular, the propagation angle of the waves in third harmonics generation with frequency $3\omega$ is $\theta \approx 52^\circ$, generated by an incident wave with frequency $\omega =0.4 \omega_P$, which also lies in the frequency range where the surface plasmon mode exists.

Lastly we estimate the magnitude of the nonlinear effect. We consider the nonlinear susceptibility
\[\chi^{(3)}(3\omega; \omega, \omega, \omega) = \bar{\gamma} \chi^{(1)} (3\omega) \left[1 + \frac{i \bar{b}}{\omega^3}\right] [\sigma^{(1)}(\omega)]^3.
\]
In the typical frequency scale $\omega_P \sim 10^{12} s^{-1}$ where the surface plasmon emerges, the linear conductivity lies in the scale $\sigma^{(1)}(\omega) \sim \Omega^{-1} {\rm cm^{-1}}$ \cite{Ma&Ng}, and the magnitude of the linear susceptibility is of order $\chi^{(1)} (3\omega) \sim \sigma^{(1)}(3\omega)/3\omega \sim 10^{-10} \Omega^{-1} {\rm m^{-1} s }$. From Eq.~(\ref{eq:taugam}), we also find $\gamma \sim 1/(k^2_F \tau) \sim \tau^{-1}/(2 m \epsilon_F)$, where $\epsilon_F \approx 10 {\rm m eV}$ is the Fermi energy. Choosing $\tau \sim 10^{-12} s$ and $\sigma^{(1)}(\omega)/\sigma_0 \sim 1$ and $\bar{\gamma} = \gamma m^3/n^3_0 e^4 = \gamma e^2 \tau^3/\sigma_0^3$. The first term of the nonlinear susceptibility is thus of order:
\begin{eqnarray}
  \chi^{(3)}(3\omega) \sim \frac{e^2 \tau^2}{2 m \epsilon_F} \chi^{(1)} (3\omega) \left[ \frac{\sigma^{(1)}(\omega)}{\sigma_0}\right]^3 \sim 10^{-21} {\rm m}^2 / {\rm V}^2. \nonumber
\end{eqnarray}
For the second term, $g$ is a dimensionless coupling constant of order unity \cite{Lee2005}, and a simple dimension analysis gives $b_1 \sim 1/ g v^2_a E^2_c$, where $E_c$ is a microscopic energy scale for charge fluctuation. For system close to metal-insulator transition, $E_c \sim \epsilon_F$, and $v_a \sim v_F \sim 10^{-4}c$. Thus, as a rough estimation, the second term is of order
\begin{eqnarray}
  \chi^{(3)}(3\omega) \sim b [\chi^{(1)}(\omega)]^4 \sim \frac{g^3 v^8_a}{g v^2_a \epsilon^2_F} [\chi^{(1)}(\omega)]^4 \sim 10^{-22} {\rm m}^2 / {\rm V}^2, \nonumber
\end{eqnarray}
which is smaller than the first term in the frequency scale considered. We can see the magnitude of the nonlinear susceptibility is comparable with the typical value $\chi^{(3)} \sim 10^{-20} {\rm m}^2 / {\rm V}^2$ \cite{boyd2008nonlinear}. Thus we expect the third harmonic generation process can be observed experimentally in the $U(1)$ spin liquids.

\subsection{Conclusion}
Summarizing, we study in this paper the nonlinear response of $U(1)$ spin liquids under the perturbation of a time-dependent but spatially uniform electric field within the framework of a $U(1)$ spin liquid theory. The third-order optical conductivity $\sigma^{(3)}(\omega)$ is determined where we show that it has the same low energy power-law structure as the linear optical conductivity $\sigma^{(1)}(\omega)$. We show that the nonlinear optical conductivity leads to an interesting, observable phenomenon where the surface modes with frequency $\omega$ supported by the complex linear susceptibility generates through the third-order process an EM wave with frequency $3\omega$ which can propagate inside the nonlinear medium. This excited mode can be detected in an Otto configuration.

\begin{acknowledgments}
We acknowledge support from HKRGC through grant No. 603913.
\end{acknowledgments}

\bibliography{SL}

\end{document}